\documentstyle[12pt,aaspp4]{article}

\def\ltsima{$\; \buildrel < \over \sim \;$}
\def\simlt{\lower.5ex\hbox{\ltsima}} 
\def\gtsima{$\; \buildrel > \over \sim \;$}
\def\simgt{\lower.5ex\hbox{\gtsima}} 

\begin{document}
\input{psfig.tex}

\title{Multiwavelength Monitoring of the BL Lacertae Object PKS~2155--304
in May 1994. III. Probing the Inner Jet through Multiwavelength Correlations}
\date{March 5, 1997}

\author{C. Megan Urry\altaffilmark{1}, A. Treves\altaffilmark{2,3}, 
L. Maraschi\altaffilmark{4,5}, H. Marshall\altaffilmark{6}, 
T. Kii\altaffilmark{7},
G. Madejski\altaffilmark{8},}
\author{S. Penton\altaffilmark{9},
J. E. Pesce\altaffilmark{1},
E. Pian\altaffilmark{1},
A. Celotti\altaffilmark{2},
R. Fujimoto\altaffilmark{7},
F. Makino\altaffilmark{7},
C. Otani\altaffilmark{7},}
\author{R. M. Sambruna\altaffilmark{8},
K. Sasaki\altaffilmark{7},
J. M. Shull\altaffilmark{9},
P. Smith\altaffilmark{10},
T. Takahashi\altaffilmark{7},
M. Tashiro\altaffilmark{11}}

\altaffiltext{1}{Space Telescope Science Institute, 3700 San Martin Drive, 
       Baltimore, Maryland, 21218\\Electronic mail: cmu@stsci.edu}
\altaffiltext{2}{SISSA/ISAS International School for Advanced Studies, 
	Trieste, Italy}
\altaffiltext{3}{Department of Physics, University of Milan at Como, 
	Via Lucini, I-22100 Como, Italy}
\altaffiltext{4}{Osservatorio Astronomico di Brera, via Brera 28, I-20121
	Milan, Italy} 
\altaffiltext{5}{Department of Physics, University of Milan, via Celoria 
	16, I-20133 Milan, Italy} 
\altaffiltext{6}{Eureka Scientific, Inc., 2452 Delmer St., Suite 100, 
	Oakland, CA 94602}
\altaffiltext{7}{Institute for Space and Astronautical Science, 3-1-1 
	Yoshinodai, Sagamihara, Kanagawa 229, Japan}
\altaffiltext{8}{Laboratory for High Energy Astrophysics, Code 666, Goddard 
	Space Flight Center, Greenbelt MD 20771}
\altaffiltext{9}{Joint Institute for Laboratory Astrophysics, University 
	of Colorado, Campus Box 440, Boulder CO 80309-0440}
\altaffiltext{10}{Steward Observatory, University of Arizona, Tucson AZ 85721}
\altaffiltext{11}{Department of Physics, School of Science, University of 
	Tokyo, Bunkyo-ku, Tokyo 113, Japan}

\setcounter{footnote}{0}

\begin{abstract}

In May 1994 the BL Lac object PKS~2155--304 was observed continuously
for $\sim$10 days with IUE and EUVE and for 2 days with ASCA, as well 
as with ROSAT and with ground-based radio, infrared, and optical telescopes.
The light curves show a well-defined X-ray flare followed by a broader,
lower amplitude extreme ultraviolet (EUV) flare $\sim$1~day later 
and a broad, low-amplitude
UV flare $\sim$2~days later. X-ray fluxes obtained at three
well separated times the preceding week indicate at least one previous
flare of comparable amplitude 
or perhaps ongoing stochastic X-ray variations, and additional rapid
variability was seen at the beginning of the IUE observation, when
extremely sharp changes in UV flux occurred.
The X-ray flux observed with ASCA flared by a factor of $\sim2$
in $\sim1/2$~day and decayed roughly as fast. In contrast, the
subsequent UV flare had an amplitude of only $\sim35$\% and 
lasted longer than 2~days.

Assuming the X-ray, EUV, and UV events are associated, the lags,
the decrease of amplitude with wavelength, and the broadening of
the temporal profile with wavelength are all qualitatively as expected for
synchrotron emission from an inhomogeneous relativistic jet.
Due to the high quality of the data, we can rule out that the
observed flares were caused by either a Fermi-type shock acceleration
event or a pair cascade in a homogeneous synchrotron-emitting region.
A homogeneous region is still possible if there was an instantaneous 
($t <<$hours) injection of high energy electrons that emit first at 
X-ray energies. Alternatively, the data are consistent with
a compression wave or other disturbance crossing a region with stratified 
particle energy distributions.
This kind of situation is expected to occur behind a shock front
and/or in an inhomogeneous jet. The present light curves
are in sharp contrast to the multiwavelength variability observed
in November 1991, when the amplitude was
wavelength independent and the UV lagged the X-rays by less than $\sim3$~hours.
This means that the origin of rapid
multiwavelength variability in this blazar is complex, involving at least two 
different modes.

\end{abstract}


\section{Introduction}

Multiwavelength observations have established that the continuum emission 
from BL Lac objects is almost certainly
produced by high energy particles in a relativistic jet (Bregman, Maraschi,
\& Urry 1987). Furthermore, the
recent detection of high-energy gamma-rays from many blazars confirms
that the emitted radiation is relativistically beamed along the jet 
axis (Maraschi, Ghisellini, \& Celotti 1992; Dermer \&
Schlickeiser 1993).
How the plasma is accelerated to relativistic
bulk velocity and how radiating particles within the plasma
are accelerated to high energies remain essentially unknown.

The multiwavelength spectra of BL Lacs are remarkably smooth and 
steepen
progressively from radio to X-ray wavelengths, with the emitted power 
per decade peaking between $10^{13}$ and $10^{16}$~Hz
(Giommi et al. 1995; Sambruna, Maraschi, \& Urry 1996).
The primary mechanism for this emission is most likely synchrotron 
radiation.
There is a separate, harder component peaking near (or in some cases 
well above) 1~GeV (von Montigny et al. 1995), 
which is probably due to inverse-Compton scattering
of lower energy photons, either synchrotron photons within the jet
or other UV photons external to the jet 
(e.g., Maraschi, Ghisellini, \& Celotti 1994,
and references therein).

One of the key problems in interpreting blazar spectra is explaining
the gradual steepening of the synchrotron power law.
One possibility is that the emission comes from a homogeneous region,
in which case the increasing effectiveness of energy losses
at higher energies can steepen the spectrum.
Alternatively, the particle spectra and other physical quantities may
vary along the jet (i.e., the jet is inhomogeneous) so that progressively
more extended regions in the jet effectively emit at increasing
wavelengths.
Either possibility is consistent with single-epoch multiwavelength 
spectra.

Spectral variability at X-ray wavelengths is common, with
spectra typically hardening with increasing intensity (Sambruna
et al. 1994, and references therein).
Multiwavelength variability can give strong clues
about the processes of particle injection acceleration and diffusion
in the emission region(s) and/or about the relative location of
emission regions for different spectral bands. That is, 
multi-epoch
multiwavelength data do constrain (and perhaps now over-constrain) current
blazar models.

Accordingly, we carried out comprehensive multiwavelength monitoring 
of
the BL Lac object PKS~2155--304 in November 1991
(Edelson et al. 1995; Smith et al. 1992; Urry et al. 1993; Brinkmann et al. 
1994;
Courvoisier et al. 1995). PKS~2155--304 is the brightest blazar at UV
wavelengths and is also very bright in the X-ray and optical, making 
it
the best candidate for monitoring at short wavelengths.
The 3.5 days of overlapping ROSAT and IUE data from November 1991
showed very rapid, highly correlated variability, with a secular
decrease superposed with multiple peaks 
of roughly equal amplitude in optical, UV and X-ray bands,
and with a possible 2-3 hour delay of the UV with respect to X-ray
light curve (Edelson et al. 1995).
The rapidity of the variations, the lack of dependence
of the variability amplitude on wavelength, and the short lag were
somewhat unexpected for synchrotron emission from an inhomogeneous
jet (e.g., Celotti, Maraschi, \& Treves 1991);
the repetitive peaks across the intensity decline were also 
intriguing.

We therefore planned a second, longer monitoring campaign with even 
more
extensive wavelength coverage, involving IUE, EUVE, ROSAT, and ASCA.
The observations were designed to resolve the fastest time scale
variability, to follow the light curve (at high temporal resolution)
considerably longer than in 1991, and
to improve the measurement of the UV-X-ray cross-correlation.
This second campaign took place in May 1994.
PKS~2155--304 was observed with IUE and EUVE for 10 and 9 days, respectively,
and for 2 days with ASCA.
A few ROSAT observations were also made just prior to
the ASCA pointing.

Here we present the multiwavelength results from the May 1994 campaign.
The observations and resulting light curves
are described briefly in \S~2, and a multiwavelength cross-correlation
analysis is presented in \S~3. In \S~4 we discuss possible interpretations
of the multiwavelength results. Full results in each wavelength range
and details of the data reduction and analyses 
are presented in separate papers (UV, Pian et al. 1997; EUV,
Marshall et al. 1997; X-ray, Kii et al. 1997; ground-based radio-optical, 
Pesce et al. 1997).
Conclusions are summarized in \S 5.

\section{Observations and Results}

In the following we describe briefly the 1994 multiwavelength observations of 
PKS~2155--304, particularly those made with high time resolution in the 
UV, extreme UV, and X-ray energy bands. We refer to the 
above quoted papers for complete presentations. Dates are reported in UT 
throughout the text, and both in UT and MJD ($\equiv$JD -- 2,440,000)
in the figures, taking into account that MJD 9487.5 corresponds 
to 1994 May 15 00:00 UT.

\subsection{Ultraviolet Light Curves}

The UV data were taken with IUE between 15 and 
25 May 1994 
with only a few short gaps due to Earth occultation. The LWP 
(long-wavelength) and SWP (short-wavelength) cameras were exposed 
alternately, with typical exposure times of 25 min and 55 min, 
respectively, in phase with the 96-minute ROSAT orbital period.
Spectra were extracted using TOMSIPS 
(Ayres 1993; Ayres et al. 1995), a modified version of the Signal-Weighted
Extraction Technique (Kinney, Bohlin, \& Neill 1991) adapted 
for the IUE Final Archive. 
Spectra were dereddened assuming $A_V = 0.1$~mag and then fitted with 
power laws separately in the 
LWP (2100 - 2800 \AA) and SWP (1230 - 1950 \AA) wavelength ranges. 
Light curves were constructed from the fitted fluxes at 2800~\AA\ and 
1400 ~\AA; the LWP fluxes have larger uncertainties because of the 
smaller effective range of the fits, given the noise below $\sim2400$~\AA\ 
and the scattered light above $\sim2800$~\AA. The mean fluxes for 
the IUE observations were 
(7 $\pm$ 1) $\times 10^{-14}$~ergs~cm$^{-2}$~s$^{-1}$~\AA$^{-1}$ for the
LWP and (15 $\pm$ 2) $\times 10^{-14}$~ergs~cm$^{-2}$~s$^{-1}$~\AA$^{-1}$ 
for the SWP. 
Further information about the IUE results is given by Pian et al. (1997).

In Figure~1 
\begin{figure}
\centerline{\psfig{figure=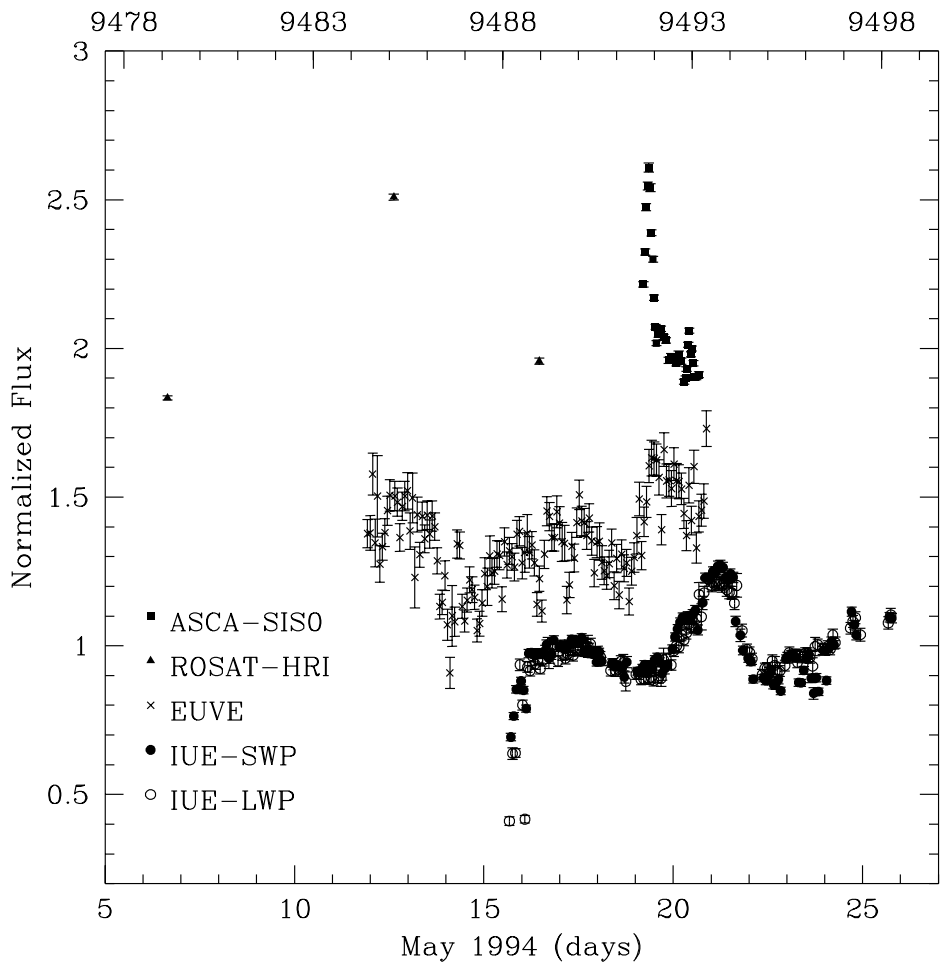,height=5.2in}}
\caption{Multiwavelength light curves of PKS~2155--304
from the May 1994 campaign.
The ASCA, ROSAT, and EUVE fluxes are averages over one orbit.
Each light curve has been normalized to its mean intensity; for the IUE data, 
this was computed after discarding the first 6 flux points from 
both SWP and LWP light curves. For clarity the EUV and X-ray light curves 
have been shifted vertically by 0.4 and 1.1, respectively.
Modified Julian dates are reported on the upper axis 
(MJD 9487.5 = 1994 May 15, 00 UT).}
\end{figure}
we show the combined SWP/LWP light curves, created by 
normalizing separately to the individual mean fluxes; the effective time
resolution of the combined light curve is $\sim48$~minutes. The two light 
curves are very similar, with a prominent flare centered on 21 May. The
amplitude of this flare is $\sim 35$\% in both cameras 
(for amplitude defined as $\Delta F / F_{initial}$) and the duration 
(FWHM) is $\sim2$~days. The structure of the flare is also similar in
both cameras, with comparable rise and decay times and 
with a flat plateau at the peak, but with a small shoulder during the 
rise. The spectral index in the short, long and combined wavelength
ranges is remarkably constant across the flare.
Some smaller amplitude variability (5-10\%) is also seen in both 
cameras throughout the observation. 

In addition, there is a dramatic event at the very beginning of the 
observation: a fast rise, followed by a dip, and another fast rise to 
the mean flux level. These initial flares represent the fastest 
ever seen at UV wavelengths in an extragalactic object. 
The doubling time scale,
defined as $\tau_D \equiv (F_{initial}/\Delta F)\Delta t$,
is $\sim 1$~hr for the LWP flux, 
comparable to the time scale of the fastest variations 
seen at X-ray and gamma-ray wavelengths.
(For the fast UV and X-ray variations discussed here, the 
e-folding times are similar.)
The doubling time for the central flare is much longer: 
$\tau_D\sim 3$-4~days.
Throughout the observation,
the spectral index varies significantly but by a small amount
(except during the first flare, when the
variability time scale is less than the temporal resolution;
Pian et al. 1997).
 
\subsection{Extreme Ultraviolet Light Curve}
 
The EUVE satellite was pointed at PKS~2155--304 for nine days, from 11 to 
20 May 1994, with the Deep Survey Spectrometer. The primary goal
of this observation was to try a new method for measuring polarization 
in the soft X-ray band, as well as to
obtain a high quality EUV spectrum of this 
BL Lac object (K\"onigl et al. 1995; see also Fruscione et al. 1994). 
The long integration time required for a good signal-to-noise ratio 
(PKS~2155--304 is one of only a handful of extragalactic objects that 
can even be detected with EUVE) resulted in the excellent long-term 
light curve discussed here. 
Additional details and results are given by Marshall et al. (1997).

The normalized EUV light curve is shown in Figure~1. The data
are binned on the orbital time scale (96 minutes) and each bin represents
500-1000~s of exposure. All variations appear to be resolved; that is,
300-s light curves reveal no flares that are not also apparent in the
longer binned data.
Although the errors on the fluxes are larger than for the
UV light curve, the EUV light curve shows a similar structure. In 
particular, a sharp rise occurs around 19 May, with a $\sim 50$\%
 amplitude and 
$\sim$1-2~day duration, very similar to the UV flare occurring about 
a day later.
Also, the minimum on 14 May resembles (within the large 
errors) the initial event seen on 16 May with IUE. This is shown with an 
expanded 
scale in Figure~2, and with the EUV light curve shifted by 1~day, 
corresponding to the peak in the EUV-UV cross-correlation function (see \S~3).
\begin{figure}
\centerline{\psfig{figure=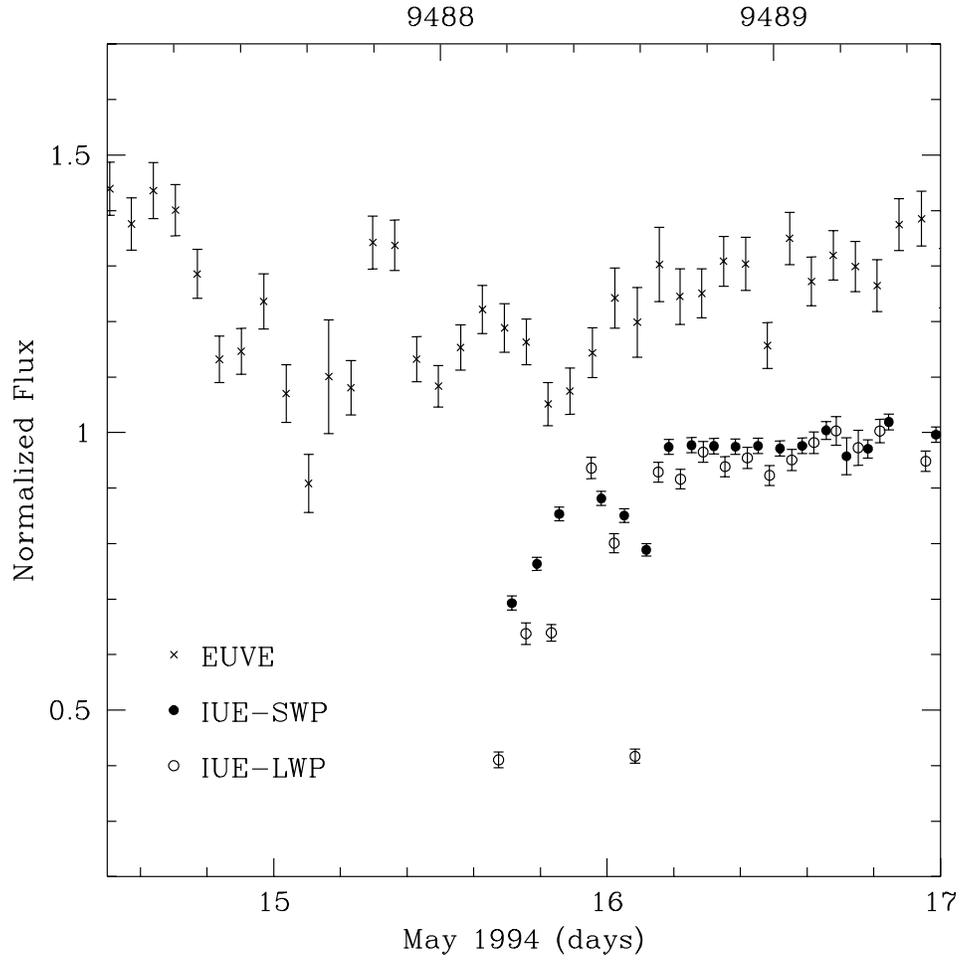,height=5.2in}}
\caption{Expanded view of part of the normalized EUVE and IUE light 
curves, with the former shifted forward by its lead time, 1~day, and upward
by 0.35, for clarity.
Although the EUVE data are noisy, variations similar to the initial UV
flares can be seen, although they are much smaller in amplitude and not
at the same lag as the central flare.
Modified Julian dates are reported on the upper axis.
}
\end{figure}

\subsection{X-Ray Light Curve}

The ASCA satellite was pointed at PKS~2155--304 for two days from 
4:30 UT on 19 May to 7:55 UT on 21 May. The goal of the
observation was to study the energy dependence of fast
X-ray variability and to follow the evolution of the X-ray spectrum 
over the wide energy band of 0.5-8 keV; these
issues are discussed in detail by Kii et al. (1997). 
Data were accumulated with the two SIS and two GIS 
focal plane detectors, which have overlapping but
slightly different energy ranges (Tanaka, Inoue, \& Holt 1994).

The results presented here are from the SIS detectors, which have
wider energy coverage and better energy resolution than the GIS
(which gives equivalent results in any case; Kii et al. 1997).
The SIS were operated in 1-CCD faint mode for
high-bit-rate data and in 1-CCD bright mode for medium bit-rate
data. The data extraction followed the standard
recommendations of the ASCA instrumental teams. Since the count rate
of the background and its fluctuations are negligible compared with
the count rate from PKS~2155--304, we performed no background
subtraction for the timing analysis so as to avoid any artificial effects.

The normalized X-ray light curve from the ASCA SIS-0 CCD is shown in Figure~1. 
A pronounced high-amplitude flare occurs at the beginning of the
observation. The observed 0.5-8 keV flux is $3.3\times10^{-10}\
\mbox{erg cm}^{-2}\mbox{ s}^{-1}$ at the peak and $1.5\times10^{-10}\
\mbox{erg cm}^{-2}\mbox{ s}^{-1}$ at the minimum. The shape of the
flare is approximately symmetric in time. 
The doubling time for the decay is $\tau_{D}\sim0.5$~day; its duration 
is less clear since the flare may well have begun before the start of 
the observation. The observed part of the flare rise is symmetric with
respect to the decay, so assuming symmetry we estimate that the flare
duration (FWHM) is $\sim 1$~day. 
Toward the end of the decay two smaller flares are visible. 
The cross correlation between the 0.5-1~keV and 2.2-8 keV bands
shows clear evidence that the hard X-rays lead the soft X-rays by 
$\sim 5\times10^{3}$ sec (Kii et al. 1997). 

We analyzed the SIS-0 spectra accumulated in each orbit, fitting
a power law model with fixed Galactic absorption 
($N_H = 1.77 \times 10^{20}$~cm$^{-2}$; Stark et al. 1992).
For the energy range 0.5-8~keV, the energy spectral index varied
from $\alpha = 1.29\pm0.02$ at the beginning of the observation 
($F_{\rm 1~keV}$ = $44.9 \pm 0.4$~$\mu$Jy), to $\alpha = 1.25\pm0.02$ 
just before the peak ($F_{\rm 1~keV}$ = $58.3 \pm 0.5$~$\mu$Jy),
to $1.80\pm0.02$ near the end ($F_{\rm 1~keV}$ = $32.4 \pm 0.4$~$\mu$Jy, 
more or less the faintest state). 
The observed spectral change is similar to previous observations of
the spectrum hardening (softening) with increasing (decreasing) intensity
(Treves et al. 1989; Sembay et al. 1993),
and is consistent with the observation that the
hard X-rays lead the soft X-rays.

PKS~2155--304 was also observed in soft X-rays with the ROSAT HRI for three
short (one orbit) periods, on 9, 12, and 16 May 1994, just before the bulk
of the multiwavelength campaign. The HRI is sensitive in the energy range
of $\sim 0.1 - 2.4$~keV, but provides little spectral information. The
data were reduced using the standard ROSAT HRI criteria. 
Source counts were extracted from a region 30~arcsec in
radius, while the background counts were extracted from an annulus with 
inner and outer radii of 3 and 5~arcmin; the background level is
typically 0.3\% of the source counts. 
No other point sources at a level $> 1$\% of the intensity of PKS~2155--304 
were present in the HRI field of view. 

Examination of the resulting light curve (Fig.~1)
reveals large amplitude variability from one ROSAT observation to
the next, with count rates of $5.12 \pm 0.05$, $9.83 \pm 0.07$, and
$5.98 \pm 0.07$ counts/sec, chronologically. 
There is no significant variability within each 2000-second observation;
for time intervals between 400~s and 2000~s, there is less than
$<3.1$\% peak-to-peak variability 
(the statistical uncertainty in each bin is less than 2.5\%). 
This is close to the systematic limit due to the residual uncalibrated
non-uniformities of the HRI; while the estimate of the
relative variability amplitude on time scales of 400~s or longer is
probably reliable at the $\sim 2$\% level, systematic effects due to the
wobble of the spacecraft every 400~s and 
non-uniformities in the photocathode of the HRI 
limit the determination of the variability
amplitude to an estimated 5\% (S. Snowden, priv. comm.).
The resulting background-subtracted count rates were converted to flux
using an absorbed power law model with the power law (energy) index of 
$\alpha =1.5$, as measured by ASCA (Kii et al. 1997),
and the Galactic column density 
$N_{\rm H} = 1.36 \times 10^{20}$~cm$^{-2}$ (Lockman \& Savage 1995).
The resulting unabsorbed 1~keV flux densities
are, $24.6 \pm 0.3$~$\mu$Jy, $47.2 \pm 0.4$~$\mu$Jy, and 
$28.7 \pm 0.4$~$\mu$Jy.

While the ROSAT sampling is modest, the HRI light curve clearly indicates
large amplitude variability in the soft X-ray range in the first weeks of
May 1994 on a time scale of several days, with an absence of such variability
on time scales of 400-2000~s. Although no variations were resolved with 
ROSAT, the HRI data suggest that the flare seen with ASCA 
was not an isolated event.

\section{Comparison of Light Curves in Different Bands}

\subsection{UV, EUV, and X-Ray Correlations}

The most striking feature of the multiwavelength light curves is the 
well-defined large-amplitude flare seen with ASCA.
The X-ray coverage is unfortunately very limited and the 
longer wavelength light curves do not overlap completely,
so we cannot be certain, even for this extensive data set, that
the X-ray flare on 19 May corresponds to the UV flare on 21 May.
In making this identification, the EUVE data are critical.
The wavelength range of EUVE (60-90\AA, with effective wavelength $\sim75$\AA)
is close to the geometric mean of the ASCA and IUE ranges, and the
intermediate properties of the EUVE flare (the time at which the flare
is centered, its amplitude, and its duration) support our assumption
that the correlations and delays derived below represent the
real propagation of a flare from short to long wavelengths
rather than the coincidental alignment of unassociated events.

We computed formal cross-correlations using the Discrete Correlation
Function (Edelson \& Krolik 1988; with modified normalization given by
Krolik, private communication). This method samples time intervals
for which data in both light curves exist, without interpolation. 
The input light curves were the SIS0 light curve binned at 96~min 
to improve the statistics, 
the EUVE light curve binned also at 96~min, and
the combined normalized IUE SWP and LWP light curves at 
approximately 48~min resolution.

The resulting cross-correlations of the ASCA-IUE and EUVE-IUE 
light curves are shown in Figures 3a and 3b, and the lags are summarized
in Table~1.
(The cross-correlation of the EUVE and ASCA light curves does not 
yield useful results due to the very short overlap.) 
The UV light curve is strongly correlated with each of the other two, 
with the UV lagging the X-ray by $\sim$2~days and 
lagging the EUV by $\sim1$~day.
In the latter case, not only do the broad flares line up with these delays,
but also the rapid increase at the beginning of the IUE observation
(15 May)
could correspond (within $\sim1/2$ day) to dips in the EUVE 
light curve (Fig.~2). 
Note that this UV flare is undersampled even with the LWP, which represents
the shortest integration time (the SWP time bins are twice
as long). Intrinsically rapid variations 
integrated over longer time periods will have lower apparent amplitudes,
as observed. (This does not apply to the flare of May 19-21, which
is well resolved.)

\begin{figure}
\centerline{\psfig{figure=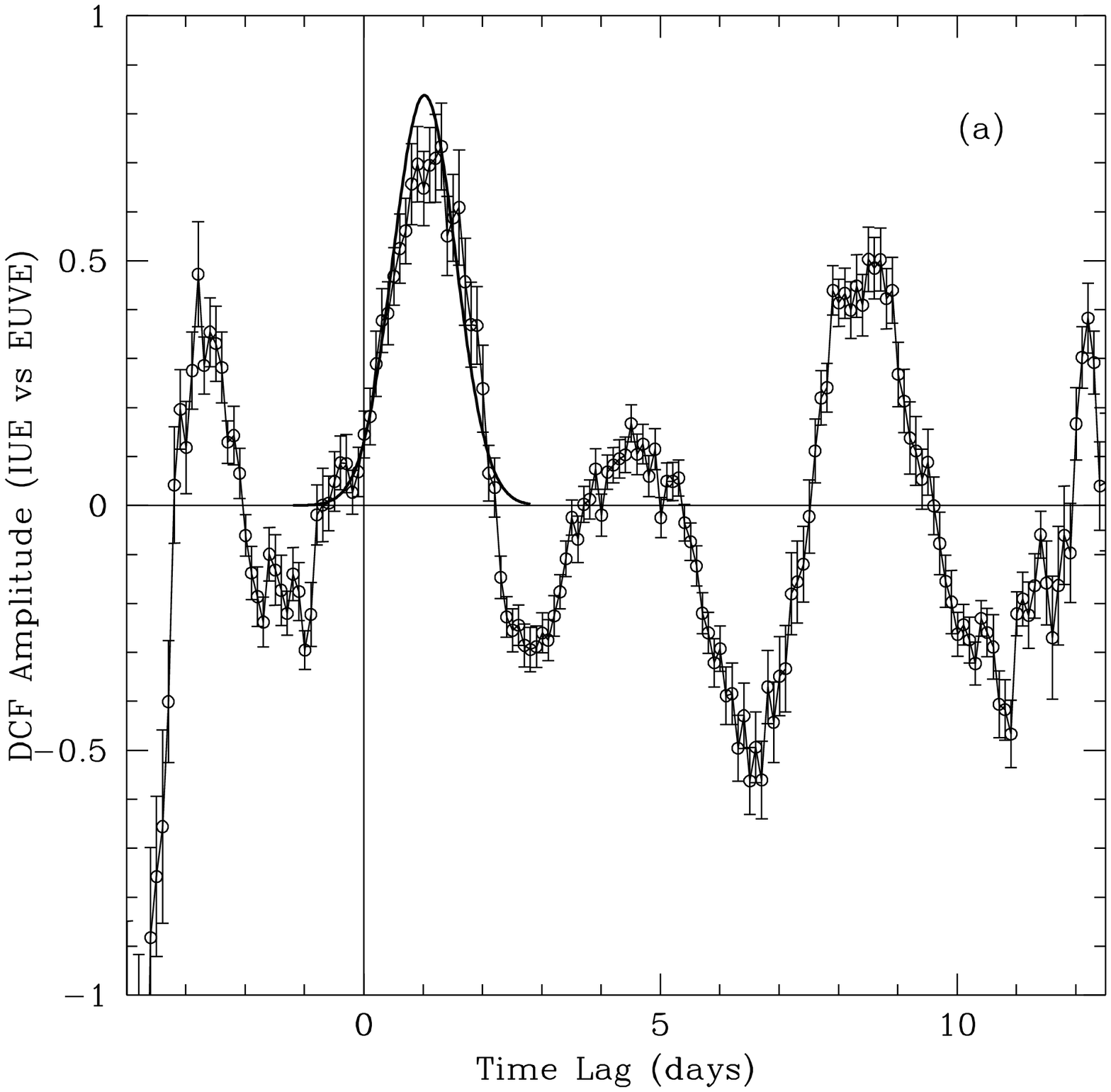,height=3.2in}}
\centerline{\psfig{figure=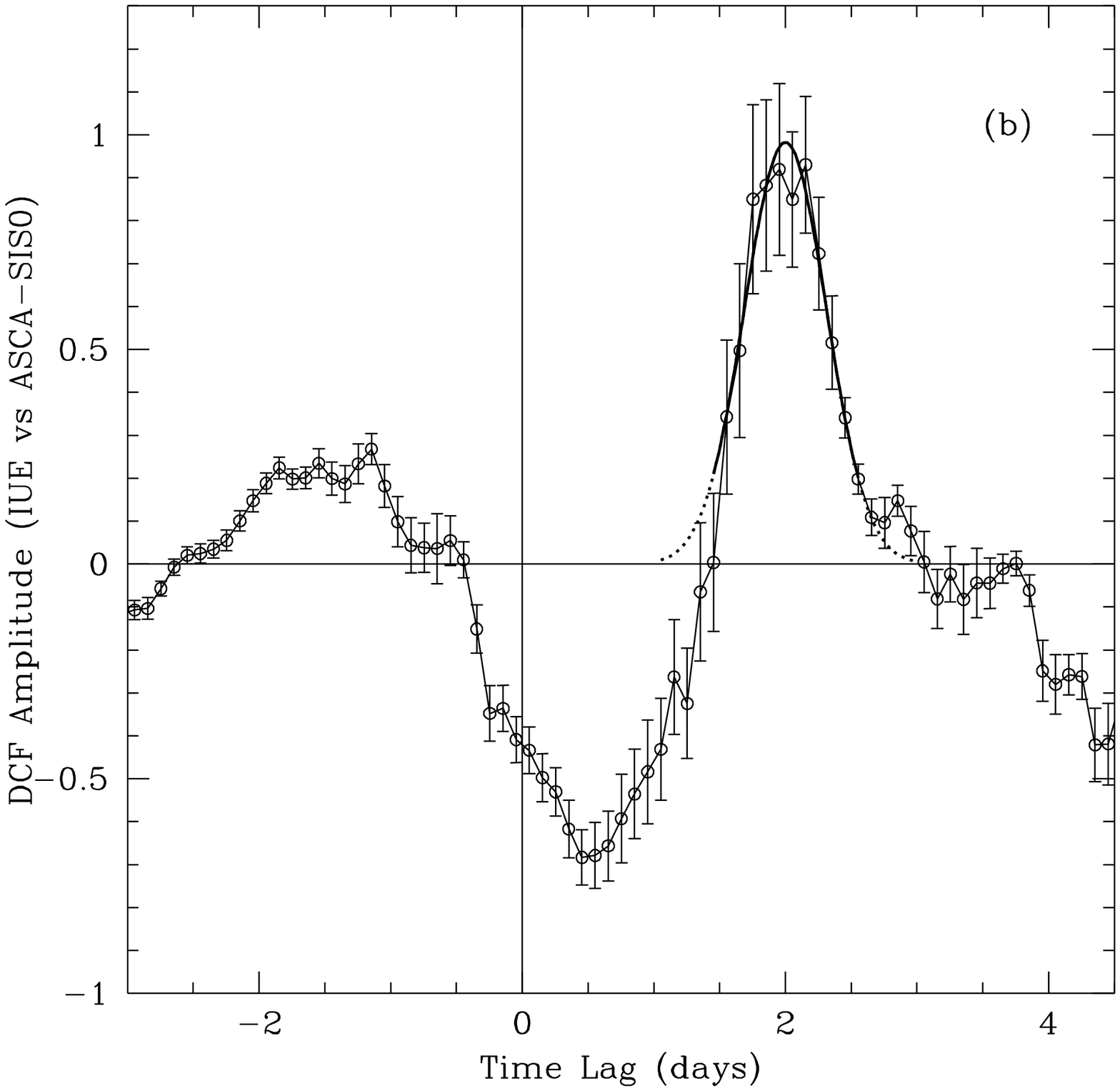,height=3.2in}}
\caption
{Cross-correlation amplitudes obtained with the DCF method.
{\it (a)} 
EUVE flux leads IUE flux by $\sim 1$~day.
{\it (b)} 
ASCA-SIS0 flux leads IUE flux by $\sim 2$~days.
Temporal lags were determined with Gaussian fits to the peaks of the
DCFs (thick solid line); dotted regions in {\it (b)} were not included in
the fit to the ASCA-IUE DCF.
}
\end{figure}

The amplitudes and shapes of the flares differ in the various wavelength
bands, as summarized in Table~1.
The flare is sharpest at X-ray wavelengths, lasting no longer than
a day, and is symmetric, with
$\tau_D \sim 0.5$~day 
(the e-folding times are similar). 
At UV wavelengths, the rise is clearly slower, with some
possible structure --- the rise and fall times are $\tau_D \sim 4$~days
and $\tau_D \sim 2.5$~days, respectively --- and the flare lasts 
2.5~days.
(Given the uncertainties, the UV light curve could represent a single flare
with a flat top or two separate flares. In the analysis below, we assume
a single flare for simplicity.)
The increase in X-ray flux is
roughly a factor of two, while the UV increase is $\sim 35$\% in amplitude.
The uncertainties in the EUV light curve are larger due to the low count
rates for PKS~2155--304 (as for any extragalactic object), 
so the flare shape is not well determined.
The flare amplitude is clearly lower than for the X-ray and more like the UV,
with a value near $\sim50$\%. The EUV rise time is $\tau_D\sim1$~day, 
faster than in the UV; the decay time is less well determined to be
$\sim2$~days but within the uncertainties could be as long as 5~days.
The duration of the
EUV flare appears to be roughly 1.5~days but because the EUVE
observation ends at that point, it could be longer.

\subsection{Comparison of Optical Flux and Polarization to UV Light Curves}

In general the optical light curves are less well sampled than the
shorter wavelength light curves, but
they have much longer time coverage (Pesce et al. 1997). 
The comparison of the V-band and UV light curves in Figure~4 
makes clear that the two total flux curves could have comparable time scales 
and amplitudes. The coverage was too sparse to associate any variations 
directly with the rapid initial UV flare on May 16, but
a few days later, an increase of $\sim 0.3$~mag ($\sim30$\%) 
is seen in both the $V$ and $R$ bands,
corresponding almost exactly to the UV flare of May 21 with
little or no lag ($\simlt 1$~day). Toward the end of the campaign, on May 25,
the {\it BVRI} fluxes show a slow increase along with the UV,
with the amplitude increasing toward the blue ($\Delta $m$ \sim 0.01$~mag
for $R$ and $I$, $\Delta $m$ \sim 0.1-0.2$~mag for $V$ and $B$). Thus the
optical and UV fluxes are generally well correlated, as observed previously.
\begin{figure}
\centerline{\psfig{figure=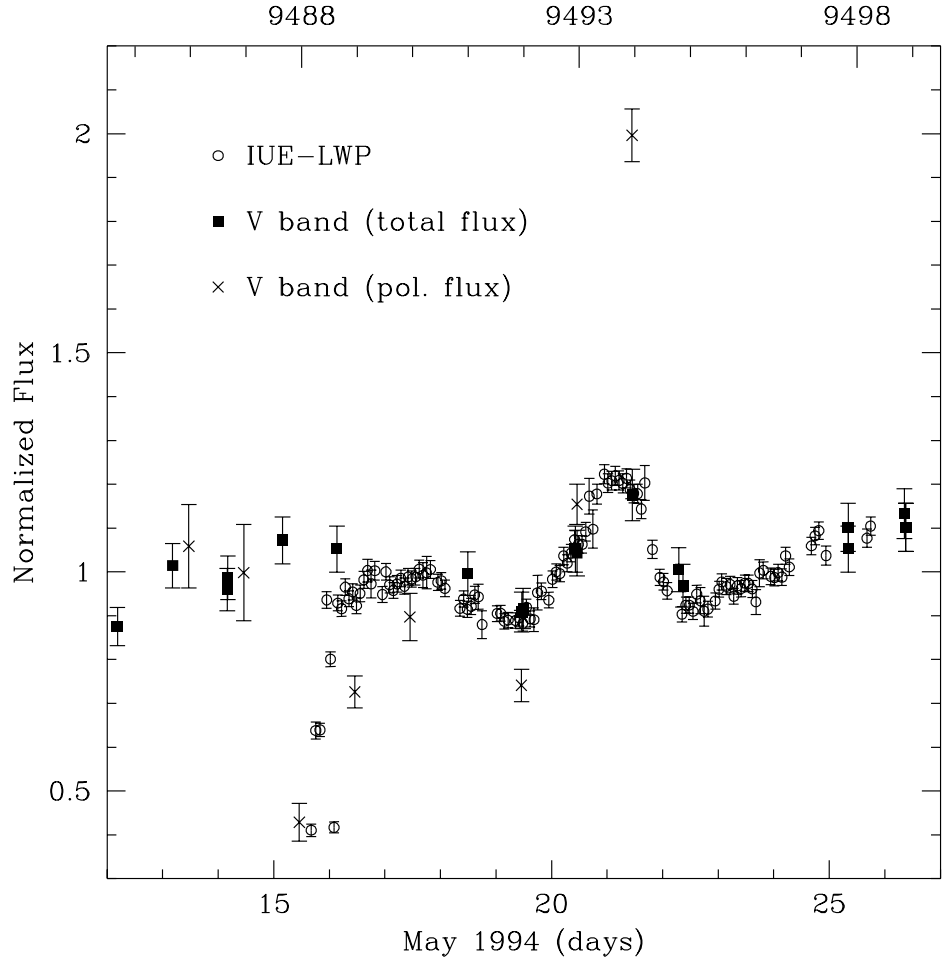,height=5.2in}}
\caption{Comparison of UV and V-band light curves of PKS~2155--304 in 
May 1994, each 
normalized to its mean. The largest variation is seen in polarized flux,
while in total V-band flux the variations are very similar to the UV
light curve. No lags were detected with respect to the UV, 
in part because of the sparse sampling of the optical data; the
peaks in the UV flux and the V-band polarization occur within $\sim 1$~day.
Modified Julian dates are reported on the upper axis.
}
\end{figure}

The variability in polarized V-band flux is much larger than in unpolarized 
flux (Fig.~4). Although the sampling is still sparse, the polarized flux dips 
dramatically when the UV flux is low during the initial flare. Later,
a strong flare in polarized V-band flux occurs at roughly the same time
as the central UV/EUV/X-ray flare, peaking perhaps one day after the
peak of the X-ray flare and the onset of the EUVE flare, and
at approximately the same time as the UV flare. 
The coincidence of these two features,
the initial and central flares, suggests a very close connection between
the polarized optical flux and the UV flux. If so, we can 
infer that the initial UV variability was indeed a dip from the
mean intensity level, rather than a flare from a lower state. 

\subsection{Broad-Band Spectral Energy Distributions}

\begin{figure}
\centerline{\psfig{figure=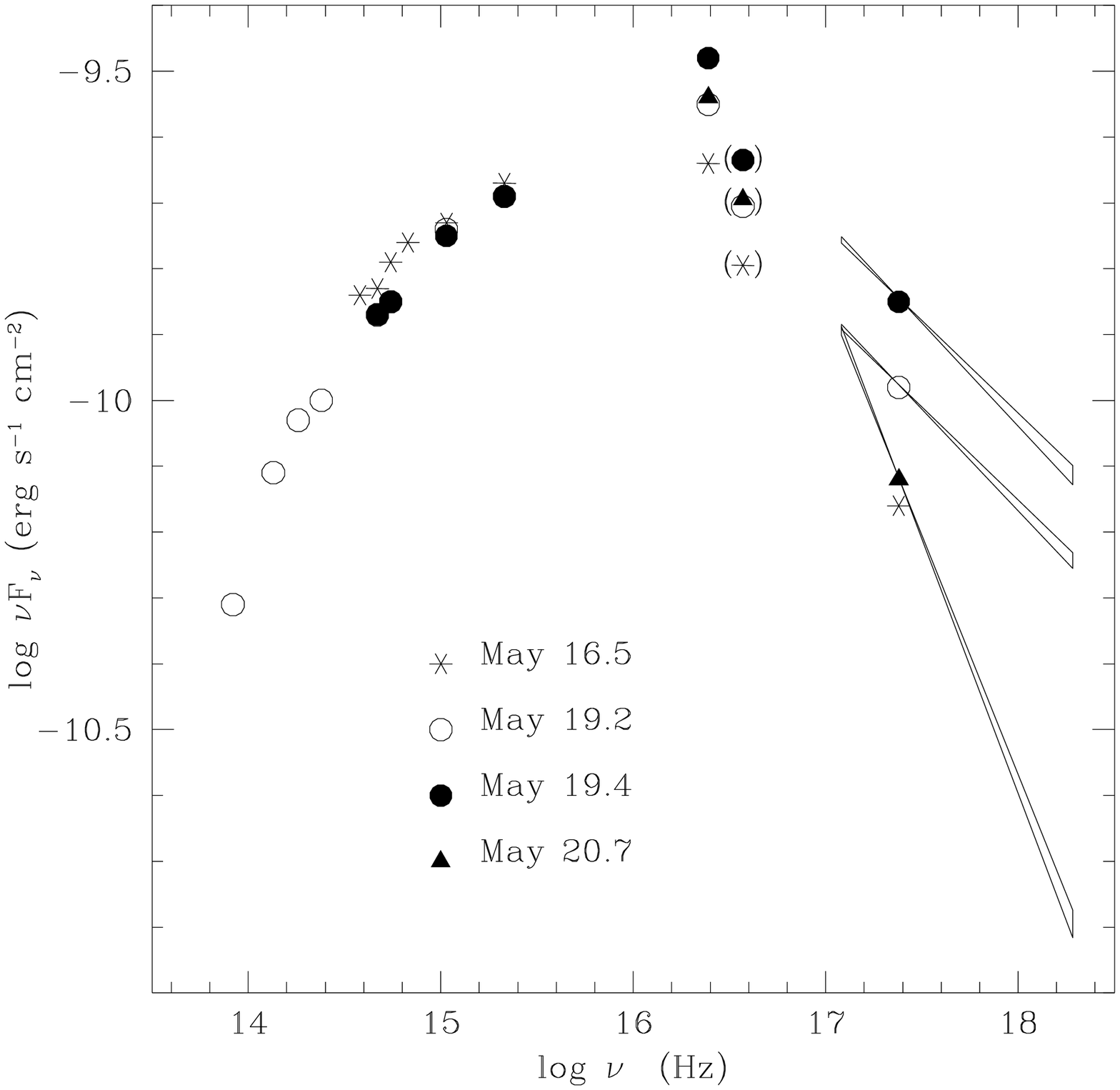,height=5.2in}}
\caption{Multiwavelength radio through X-ray spectral energy 
distributions of PKS~2155--304 at
four epochs before and through the central flare.
The X-ray spectrum clearly softens with decreasing intensity, as
the energy appearing initially at X-ray energies is manifested, with
approximately the same power, at EUV and then UV wavelengths.
The EUV fluxes are strongly dependent on the assumed $N_H$ along
the line-of-sight; points are plotted for the best Galactic value,
$N_H = 1.36 \times 10^{20}$~cm$^{-2}$ (Lockman \& Savage 1995), and in 
parenthesis for a value
only $0.1 \times 10^{20}$~cm$^{-2}$ lower, a typical uncertainty on
arcminute scales (Elvis, Lockman, \& Wilkes 1989).
}
\end{figure}
As a complementary view to the light curves in Figure~1, we show in Figure~5
four simultaneous broad-band spectral energy distributions (SEDs) which
describe the spectral evolution during the correlated multiwavelength flare.
The chosen epochs are 
(1) in the ``quiescent'' period, on 16.5 May;
(2) at the beginning of the X-ray flare, on 19.2 May; 
(3) at the peak of the X-ray flare, on 19.4 May; and
(4) at the peak of the UV flare, on 20.7 May 
(using the closest X-ray data, from 12~hours earlier).
The EUV fluxes are strongly dependent on the assumed $N_H$ along
the line-of-sight, so we have plotted points for the best Galactic value,
$N_H = 1.36 \times 10^{20}$~cm$^{-2}$ (Lockman \& Savage 1995), and 
in parenthesis for a value only $0.1 \times 10^{20}$~cm$^{-2}$ lower,
a typical uncertainty on arcminute scales (Elvis, Lockman, \& Wilkes 1989).
Figure~5 shows that already at the beginning of the X-ray flare
the X-ray spectrum hardens, grows in intensity up to the peak,
and steepens going back to the pre-flare intensity while the UV rises.
The spectral evolution of the flare shows that the flare power
is first injected at high energy and then migrates to lower energies. 
Note that the peak powers emitted at X-ray and (later) at UV energies
are of the same order.

\section{Discussion}

\subsection{General Considerations for the Synchrotron Model}

In PKS~2155--304 the emission mechanism from the UV through the 
medium energy X-ray band is synchrotron radiation, 
according to many multiwavelength spectral analyses 
(Edelson et al. 1995, and 
references therein). This is supported by optical and UV variability
(Urry et al. 1993), and particularly by the 
large and variable polarization at those wavelengths (Smith et al. 1992).
The ASCA spectrum, with
energy spectral index $\alpha\sim 1.3-1.8$ (Kii et al. 1997), 
is steeper than the UV ($\alpha\sim1$; Pian et al. 1997), as expected 
for the high energy tail of the synchrotron spectrum. It is apparent
from Figure~5 that the peak luminosity in the synchrotron component of
the spectral energy distribution of PKS~2155--304 falls in the EUV band,
corresponding to the definition of a ``high-frequency-peaked'' BL Lac
object (HBL; Padovani \& Giommi 1995). 

PKS~2155--304 has been detected with
the Compton GRO EGRET instrument with a flat ($\alpha = 0.7 \pm 0.3$)
spectrum (Vestrand, Stacy, \& Sreekumar 1996) 
and possibly with OSSE (McNaron-Brown et al. 1995), implying the presence of a
separate hard component. This component does not contribute significantly
to the ASCA spectrum since no indication of high-energy flattening 
is found.
The OSSE flux greatly exceeds the extrapolations from either EGRET 
or ASCA spectra except for the short-lived flare peak; it may possibly
be contaminated by a nearby Seyfert galaxy, NGC 7172.

The spectral energy distribution and the X-ray behavior of PKS~2155--304 
closely resemble those of another BL Lac object, Mrk~421. 
In particular, in Mrk~421 the soft X-ray photons also
lag the harder ones by about two hours.
The gamma-ray emission in both objects can derive from Compton scattering of 
high energy electrons off the synchrotron photons (the SSC mechanism;
e.g., Ghisellini \& Maraschi 1996). We also note that Mrk~421 is a strong
source of TeV photons, and expect that PKS~2155--304 would be detected as
a TeV source with a southern array (Stecker et al. 1996).

Emission models of blazars require that the observed radiation is Doppler 
boosted, implying that the 
emitting plasma moves at relativistic speed.
Different emission regions contribute to the low frequency end (radio to 
millimeter) of the spectrum yielding a relatively flat overall energy 
distribution (e.g., K\"onigl 1989). 
At higher frequencies the spectrum steepens and could be produced in a 
single region (homogeneous model) or alternatively the steepening of the
continuum could be associated with further structure in the jet, in the sense
of higher frequencies being produced closer to the jet core (inhomogeneous
jet; Ghisellini, Maraschi, \& Treves 1985). 
The observed SEDs can be explained reasonably well in both cases. 
However, the homogeneous model is fully constrained if the SED is known
up to the gamma-ray range and the variability time scale is taken to
infer the size, while the inhomogeneous model has more degrees 
of freedom. In the following we discuss the implications of the present 
results within the two scenarios.

A model-independent conclusion from our data is that the high energy particles
causing the synchrotron flare cannot be produced by a stochastic acceleration 
process, like diffusive acceleration at a shock front.
Since shock acceleration is a cumulative process,
the density of the lower energy particles builds to a maximum first, 
followed by particles of higher energy,
as shown by detailed computations (Fritz \& Webb 1990; this calculation
neglected Compton losses and applied to a nonrelativistic shock wave).
If this type of acceleration process caused the observed flare in 
PKS~2155--304,
the X-ray flare would peak after the UV flare, 
contrary to what is observed.
The data suggest alternative types of acceleration mechanisms, such as
acceleration by large scale electric fields 
(e.g., Bednarek, Kirk, \& Mastichiadis 1996a,b).
The observed lag of the soft photons with respect to the hard could be
caused by particles with initially high energies 
evolving via energy losses to lower energies (in either the homogeneous
or inhomogeneous case),
or to inhomogeneous particle distributions affected by
a passing disturbance (e.g., a moving shock or compression wave), 
or to some combination of the two. 

\subsection{Homogeneous Models}

The simplest cause of a flare in a homogeneous synchrotron model 
would be a sudden uniform increase in the density of energetic electrons.
The multifrequency light curves expected from this kind of model
need to be calculated in detail, since they depend on 
(a) the time scale and spectrum of energy injection,
(b) the evolution of the particle spectrum (a balance between acceleration
and radiative or other losses), and
(c) light travel time effects.
Some spectra have been computed
for a limited set of time-dependent particle distributions
and applied to Mrk~421 
(Mastichiadis \& Kirk 1997, neglecting light travel time effects; 
Ghisellini \& Chiaberge 1997). These imply magnetic fields and Doppler
factors very similar to those derived independently from the X-ray lags. 
Similar calculations should be applied to PKS~2155--304 but are beyond the
scope of this paper.

Qualitatively, the observed variations in PKS~2155--304 --- 
largest and fastest at the shortest wavelengths --- 
follow the trend expected for the spectral evolution of relativistic
electrons suffering radiative losses.
Synchrotron (and Compton) losses are energy-dependent, with radiating
particles having lifetimes inversely proportional to frequency,
\begin{equation}
t_s = 2\times 10^4 B^{-3/2} \delta^{-1/2}{\nu_{15}}^{-1/2}~{\rm s}
\end{equation}
(Tucker 1975),
where $t_s$ and $\nu_{15}\equiv\nu / 10^{15}$~Hz are in the observer frame,
$B$ is in Gauss, and $\delta \equiv (\gamma (1-\beta \cos \theta))^{-1}$
is the Doppler factor.

A soft lag in a uniform model can only arise
if lower energy particles derive from initially more energetic ones 
after substantial energy loss. The injection of particles at high energies 
must be instantaneous across the volume (relative to the observation time
scale) rather than building from lower energies. The increase in high
energy particles must then be balanced by losses such that the X-ray pulse 
decays before the UV flare rises. The injection can not occur far above 
X-ray-emitting energies since the pulse should broaden as it moves down in 
energy.
Whether this set of constraints can be met by
a homogeneous model depends on detailed calculations not attempted for
this paper. 

Simple estimates force us to continue to consider the homogeneous
case, however. We note that the lag must be of the order
of the radiative time at the softer energy and the lags at different 
frequencies should approximately satisfy equation~(1) above 
($B$ is constant by assumption in the homogeneous case). 
In fact, the lag of 2~days at UV wavelengths,
the lag of 1~day at EUV wavelengths, and the 
lag of $5\times 10^3$~s at soft X-ray wavelengths
(all with respect to hard X-rays)
are roughly consistent with the prediction of equation~(1).

If the observed lag of the soft X-ray photons vs. the hard ones
is attributed to radiative losses, an estimate of the magnetic field follows.
For a Doppler factor $\delta \sim$ 10, 
the observed lags imply $B\sim 0.1$-0.2~G.
These values are very similar to those derived for Mrk~421 by Takahashi 
et al. (1996), which had similar X-ray lags.
For Mrk~421 a fit to the broad-band spectrum with a homogeneous
SSC model requires a Doppler factor of 18 in order not to exceed the
observed gamma-ray emission (Ghisellini, Maraschi, \& Dondi
1997; Mastichiadis \& Kirk 1997). For PKS~2155--304 the required
Doppler factor is still higher, close to $\delta \sim 30$ if the
(unmeasured) gamma-ray flux at the time of our observations was 
as bright as that measured by EGRET 6 months later; $\delta$ 
must be even larger if the actual level of gamma-ray emission was lower.

We conclude that a sudden injection of high energy particles in a uniform 
region may be a viable explanation of the observed light curves.
It does require a very high value of the Doppler beaming factor. More
detailed, time-dependent calculations are needed to check whether this
scenario is indeed possible.

It appears unlikely that the energy release occurs through a pair cascade,
since the spectra computed from
pair cascades tend to be very broad (Levinson \& Blandford 1995).
In contrast the flare fades
in X-rays before rising in the UV implying a relatively narrow energy 
distribution at a given time. 

\subsection{Inhomogeneous Models}

An alternative way to have the X-ray flare precede the UV
is that particles of different energies have different locations
along the jet axis.
(Note that at a fixed wavelength, the dominant emission region 
may still be approximately homogeneous.)

This inhomogeneity could occur 
behind a shock front,
where particles continuously accelerated at the shock fill an 
energy-dependent volume determined by radiation losses 
(a ``stratified shock''; Marscher 1996). 
In this case the lags would still be of the order
of the radiative lifetimes, so the parameter estimates
of the previous section apply. 
Also, the volume filled by particles of a given energy is related
to their radiative lifetimes so that the flare could naturally be
symmetric. The requirement on the Doppler factor could be 
somewhat relaxed since there are larger volumes for the lower energy
particles. A quantitative model should be computed to assess this case.

A different scenario is that of an inhomogeneous jet, where 
the particle energy distributions change along the jet as a result of 
a local balance between acceleration and energy loss processes.
The main difference between 
the two pictures is that in the first case particle acceleration is 
localized at the shock front while in the second it 
occurs throughout the jet volume.
In both cases, spectral steepening from UV to X-ray
is due to the diminishing volumes occupied by particles of higher energies.

A disturbance crossing a steady-state shock front or propagating along an
inhomogeneous jet
would affect the higher energy particles first and extend gradually
to the larger region occupied by the lower energy particles. 
The ``thickness'' of the disturbance 
(its length) should be comparable to
or larger than the size of the X-ray-emitting region, 
or else the amplitude of the 
X-ray flare would be smaller (and indeed not significantly advanced
with respect to the UV). 

The radiative lifetimes can be very short
in this scenario and variability time scales are associated with the 
travel time of the disturbance through the emission region.
This is comparable to the light travel time if the disturbance is
a relativistic shock wave propagating in the same direction as the
jet flow. The predicted variability for this kind of model 
(Celotti et al. 1991) agrees qualitatively with the variability 
observed in PKS~2155--304.
The duration of the pulse is related to the size of the
emission region at that wavelength, and in typical models the 
UV-emitting region is bigger than the EUV which is bigger than
the X-ray. The delay between X-ray, EUV, and UV would correspond
to the linear distance between the centers of the respective
emission regions. We note that the
observed delays are of the same magnitude as the duration of the flare, 
as expected if the disturbance moves smoothly from X-ray-
to EUV- to UV-emitting regions.

It is beyond the scope of this paper to fit specific time-dependent
inhomogeneous models to the flare evolution. We do note that 
this kind of model can have a stronger magnetic field than
the homogeneous case because it is not dictated by the observed time scales;
thus inhomogeneous models can account for the average SED,
including the gamma-rays, with a lower value of the Doppler beaming factor.

\subsection{Contrast to Variability of PKS~2155--304 in November 1991}

The light curves obtained in 4.5 days of monitoring of PKS~2155--304
in November 1991 (Edelson et al. 1995)
are substantially different from those discussed
here. The amplitude of variability in 1991, from X-ray to UV to optical,
was considerably smaller, with $\sim10$\% variations in single flares
and an overall $\sim30$\% decline in intensity, and was independent of 
wavelength. Equally striking,
the X-ray flux led the UV by no more than 2~hours.

The close correlation of optical and UV light curves in 1991 ruled out
substantial contribution from an accretion disk, and the close
correlation of all the light curves supported the notion of a
common origin for the emission, quite plausibly synchrotron emission.
However, the constraints on models that would be derived from the
1991 data would be entirely different from those implied by the 1994
data. That is, either the variability is caused by some
second mechanism or the parameters of the emitting region have
changed considerably.

In the spirit of our general discussion, 
we consider briefly the former possibility. 
Because the 1991 variability was
achromatic, which is not natural for intrinsic changes in a synchrotron 
emitter, it may be extrinsic. Specifically, it could
be caused by microlensing by stars in an intervening galaxy (Treves 
et al. 1997). There is a strong Ly-$\alpha$ absorption system at a
favorable location for lensing ($z=0.059$, Bruhweiler et al. 1993), 
roughly half-way to PKS~2155--304 ($z=0.116$; Falomo et al. 1993).
If the size of the emission region does not depend strongly on wavelength,
then microlensing can cause achromatic variations (Schneider \& Weiss 1987,
Kayser et al. 1989). The apparent relativistic motion of the jet
with respect to the lensing stars can cause time scales as short as
those observed in PKS~2155--304 (Gopal-Krishna \& Subramanian 1991).
The {\it a priori} probability of detecting a microlensing event during the 
1991 campaign was estimated by Treves et al. (1997) to be a few percent,
small but not negligible.

Alternatively, it was suggested (Urry et al. 1993) that the 1991 variability
could be associated with helical trajectories of
moving knots in a relativistic jet (Camenzind \& Krockenberger 1992), 
that is, due essentially to a change in viewing angle.
Such an explanation could not cause the strong wavelength
dependence of the variations seen in May 1994.

\section{Conclusions}

We have obtained multiwavelength light curves of unprecedented 
coverage and time resolution for any blazar. 
The emission at X-ray,
EUV, and UV wavelengths is well correlated but with significant
delays of approximately 1-2~days and the soft X-ray photons
lag the hard by about $\sim1$-2 hours. These lags
approximately satisfy a $\nu^{-1/2}$ relation, 
suggesting they are related to radiative losses.

The apparent progression of a strong X-ray
flare to longer wavelengths rules out a stochastic acceleration process
in a homogeneous volume.
It can possibly be explained by an instantaneous injection
of high energy particles near X-ray-emitting energies and subsequent energy
degradation in a homogeneous radiating synchrotron source. This injection
could be achieved by a single-stage acceleration, as with a large-scale
electric field (Bednarek et al. 1996a,b).
A radiative interpretation of the observed lags then constrains the
magnetic field to be rather low and requires a very high value of the 
Doppler beaming factor ($\delta\sim 30$) in order not to exceed 
the highest gamma-ray flux observed (not simultaneously) with EGRET.
Such a homogeneous model may be finely tuned but can not be ruled out with
present data.
The estimated physical parameters are similar to those derived with the
same assumptions from similar data for the BL Lac object Mrk~421
(Takahashi et al. 1996). 

In contrast, an inhomogeneous jet model incorporating acceleration by shocks
can represent both the broad-band spectrum and the
observed spectral evolution of the flare, 
with a higher magnetic field. The cause of the 
flare in this model is a propagating disturbance affecting different
regions at different times. 
The flatter X-ray spectra during the rise of the flare 
can be accounted for with an increase 
in the cutoff energy of the electron spectrum at the site of the 
perturbation, possibly due to shock acceleration. 
A stratified shock model provides an acceptable manifestation
of the inhomogeneous case but requires the time scales to be close 
to the radiative lifetimes of the particles as in the homogeneous model.
An energy-dependent volume does however relax somewhat the constraints on the 
Doppler beaming factor.

The different multifrequency behavior observed in 1991 indicates that 
different mechanisms of variability can operate in this source.
It is important to confirm that at least large flares have 
a qualitatively constant behavior. In particular the lags
should not change substantially in the homogeneous energy-degradation model.

\acknowledgements 

CMU, JEP, and EP acknowledge support from NASA Grants
NAG5-1034 and NAG5-2499, and EP acknowledges the support of a 
NATO-CNR Advanced Fellowship. Some of this work was completed while LM
was at STScI courtesy of the STScI Visitor Program. RMS acknowledges 
support from an NRC Research Associate Fellowship.

\newpage
\begin{center}
\begin{tabular}{lcccc}
\multicolumn{5}{c}{{\bf Table 1:} Multiwavelength Variability Time Scales$^a$}\\
&&&&\\
\hline
\hline
&&&&\\
Band & $\nu_{eff}$ & $\tau_D$ & Time Lag & Duration \\
     & (Hz)        & (days) & (days) & (days) \\
&&&&\\
\hline
&&&&\\
ASCA-SIS0 (0.5-8 keV) & $3.0\times 10^{17}$ & 0.5 & -- & 0.8 \\
ASCA-SIS0 (0.5-2 keV) & $2.0\times 10^{17}$ & --  & 0.06$^b$ & -- \\
EUVE (50-150 eV)      & $4.0\times 10^{16}$ & 1--2 & 1$^c$ & 1.5 \\
IUE (1200-2800 \AA)   & $1.5\times 10^{15}$ & 3--4& 2$^c$ & 2.5 \\
&&&&\\
\hline
\multicolumn{5}{l}{$^a$ Characteristics of central flare on 19-21 May.}\\
\multicolumn{5}{l}{$^b$ Temporal delay with respect to the SIS0 flux
in the 2--8 keV band.} \\
\multicolumn{5}{l}{$^c$ Temporal delay with respect to the full-band SIS0
flux (0.5-8 keV),}\\
\multicolumn{5}{l}{ ~~ computed from cross-correlations of the UV light
curve with the}\\
\multicolumn{5}{l}{ ~~ EUVE and with the ASCA light curves.}\\

\end{tabular}
\end{center}
\end{document}